\documentclass[final]{IEEEcsmag}

\usepackage[colorlinks,urlcolor=blue,linkcolor=blue,citecolor=blue]{hyperref}
\usepackage[T1]{fontenc}
\usepackage{upmath}
\usepackage{subcaption}
\usepackage{stfloats}

\newcounter{capabilityCounter}
\newcommand{\capability}[1]{\stepcounter{capabilityCounter} \textbf{C.\arabic{capabilityCounter} #1:}}

\usepackage{ifthen}

\newboolean{showcomments}
\setboolean{showcomments}{true} 
\ifthenelse{\boolean{showcomments}}
{\newcommand{\note}[2]{
		\fbox{\bfseries\sffamily\scriptsize#1}
		{\sf\small$\Rightarrow$\textit{#2}$\Leftarrow$}
	}
}
{\newcommand{\note}[2]{}
}

\begin{document}


\title{Efficient Logging for Blockchain Applications}

\author{C. Klinkm{\"u}ller}
\affil{CSIRO Data61}

\author{I. Weber}
\affil{TU Berlin}

\author{A. Ponomarev}
\affil{Exponential Trading}

\author{A. B. Tran}
\affil{Deputy}

\author{W. van der Aalst}
\affil{RWTH Aachen University}


\begin{abstract}
Second generation blockchain platforms, like Ethereum, can store arbitrary data and execute user-defined smart contracts. 
Due to the shared nature of blockchains, understanding the usage of blockchain-based applications and the underlying network is crucial. 
Although log analysis is a well-established means, data extraction from blockchain platforms can be highly inconvenient and slow, not least due to the absence of logging libraries. 
To close the gap, we here introduce the Ethereum Logging Framework (ELF) which is highly configurable and available as open source. 
ELF supports users (i) in generating cost-efficient logging code readily embeddable into smart contracts and (ii) in extracting log analysis data into common formats regardless of whether the code generation has been used during development. 
We provide an overview of and rationale for the framework's features, outline implementation details, and demonstrate ELF's versatility based on three case studies from the public Ethereum blockchain.
\end{abstract}

\makeatletter
\def\@maketitle{
\vspace*{3pt}

%



{\titlefont {\@title }\par \vskip 36pt}
{\@author }
\vskip 37pt
\ifvoid \abstractbox 
\else {\reset@font \box \abstractbox }
\fi
\par \vskip 0.5pt\par \addvspace {3\baselineskip}
}
\makeatother

\maketitle

\chapterinitial{Introduction}

\emph{Blockchain} technology was first proposed in 2008 as a peer-to-peer distributed ledger system for financial transactions 
\cite{Nakamoto.2008}. 
The second generation of blockchain platforms, like Ethereum \cite{Wood}, brought the capability to store arbitrary data and execute user-defined source code, so-called \emph{smart contracts}. 
When designing \emph{blockchain applications} \cite{Xu:2019:Blockchain-Book}, developers should consider the platform's shared nature, possibly with democratic governance among a network of unknown participants. 
This is particularly true for \emph{decentralized applications (dapps)}, which provide their main functionality through smart contracts \cite{Xu:2019:Blockchain-Book}.
While developers can exercise full control over their dapp's features, the control over when, where, and under what circumstances it is executed is limited.
Moreover, dapps might impact each other even when being functionally independent. 
For example, for some period of time the game CryptoKitties caused a transaction peak and thereby slowed down transaction processing for all other applications; and the infamous DAO attack eventually resulted in a split between Ethereum and Ethereum classic. 


In this situation, understanding how one's application is used and how the underlying blockchain is operated is critical, for planning improvements, adaptation, and failure analysis. 
Logs have been used for such purposes since the early days of computing \cite{Oliner:2012:CACM}. 
While all data is in principle present, given the immutable ledger, extracting it can be highly inconvenient and very slow \cite{DiCiccio+.2018}. 
As such, it is surprising that \emph{so far no logging libraries for Ethereum exist}.
We fill this gap by proposing the \emph{Ethereum Logging Framework} (ELF), a highly configurable logging framework that covers (i) generation of cost-efficient logging code to be embedded in dapps and (ii) extraction of data into formats suitable for analysis. 
Note that the latter applies to \emph{any} application, whether or not the code generation has been used during development.

This framework is based on our own prior work \cite{Klinkmuller+.2019} where we focused on extracting 
event data in the XES format for process mining.
The version proposed here supports general logging and log extraction capabilities, without a specific process focus. 
Among others, it further supports extraction of state and transaction data, formatting as textual logs or CSV, data streaming, and generalized selection criteria. 
ELF also allows users to extract data for business-level analysis, e.g., in process mining~\cite{vanDerAalst.2016}.

The remainder of the paper starts with an introduction of the framework's features. 
In this context, we discuss particularities related to logging on Ethereum and provide a rationale for the framework's features.
Subsequently, we cover implementation details and discuss selected case studies which showcase the versatile applications of the framework.
Finally, we conclude the paper with a summary.
\section{FRAMEWORK}

\begin{figure*}
\centerline{\includegraphics[width=26pc]{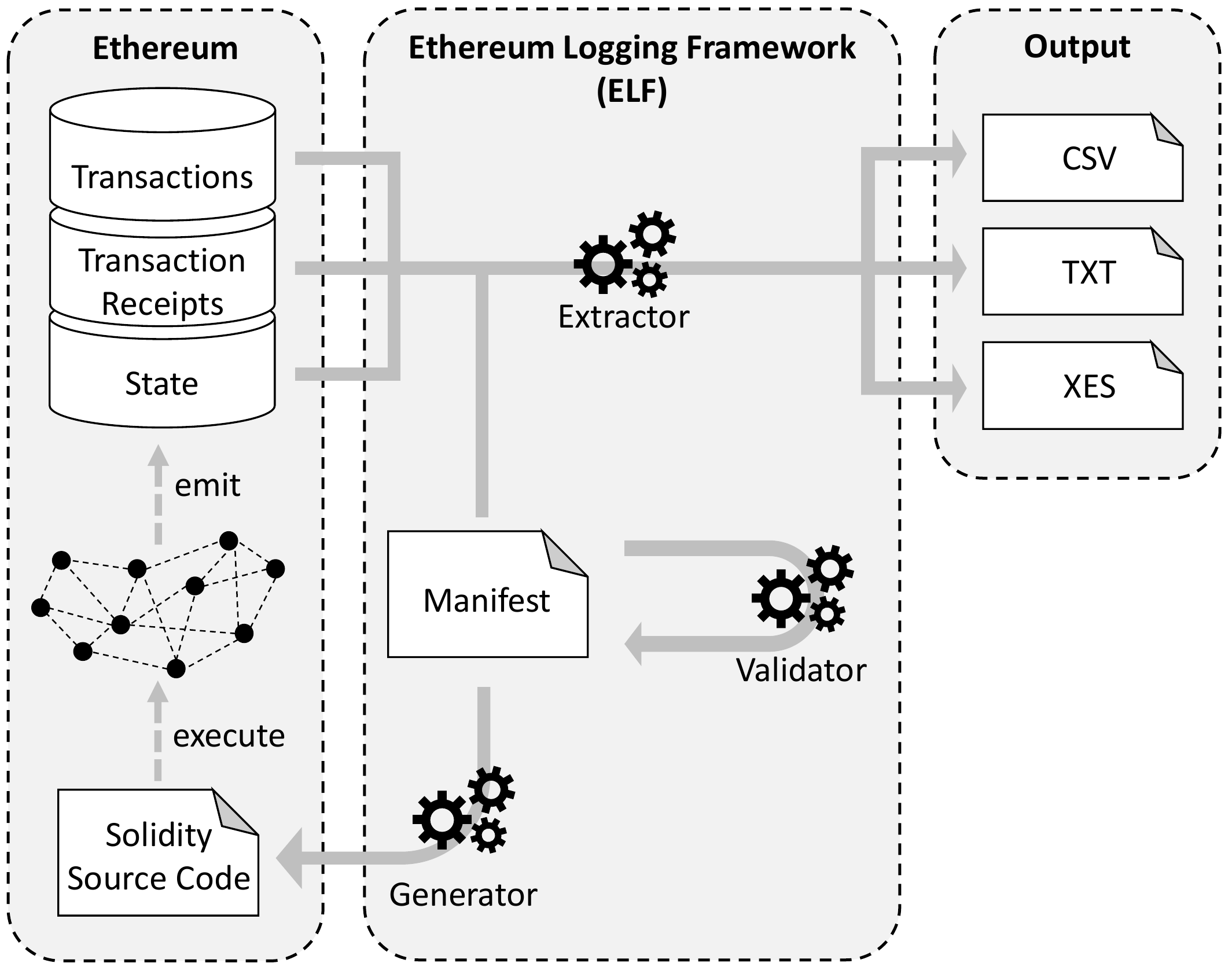}}
\caption{The components of the Ethereum Logging Framework (ELF)}
\label{fig:framework}
\end{figure*}

An overview of ELF and its components is depicted in \autoref{fig:framework}. 
At the heart of the framework is the \emph{manifest}.
It allows users to define which data in Ethereum is relevant and how it should be transformed and formatted. 
To support users in defining a manifest, the \emph{validator} syntactically and semantically analyzes the manifest for specification errors.
The \emph{extractor} relies on a given manifest to iteratively retrieve data from an Ethereum node, to subsequently transform it according to user-defined rules and to finally format the transformed data in the requested target format(s).
The \emph{generator} can derive cost efficient logging functionality from the manifest.
We provide a more detailed overview of the framework's capabilities below. 

\capability{Technological independence} 
The extractor relies on Ethereum's JSON-RPC API \cite{JSON-RPC} which is supported by the Ethereum clients Parity, Geth, and Pantheon.
The API standardized the most common types of Ethereum data queries.
An immediate consequence is that the framework can be applied to new and existing smart contracts, but also independent of the specific Ethereum node implementation and network.
Moreover, the manifest forms the only input to the extractor, validator and generator which do not require access to the source code of the dapp or smart contract.
Thus, an interpretation of dapp data (i.e., the manifest) can be shared without compromising the confidentiality of the source code. 
Note that the manifest might reference parts of the public interface of a smart contract, including log events and access to member variables.

\capability{Declarative specification}
The manifest relieves users from manually implementing an extraction pipeline. 
Instead, it enables users to declaratively configure the extraction process in a script language that hides implementation-specific details such as byte decoding, composition of API calls, transformation operations, and value formatting.

\capability{Selective data extraction}
To extract dapp related data, developers can resort to databases containing all Ethereum data.
Such databases are provided by third-parties, for example, the Ethereum dataset hosted by Google BigQuery \cite{Day+Medvedev.2018}.
Alternatively, they can be manually set up using libraries like Blockchain-ETL \cite{BlockchainETL.2019}.
When one is only interested in the data relevant to a certain dapp, this can be impracticable for several reasons.
Accessing existing databases is typically not free of charge and presumes that users trust the integrity of these databases, whereas manually setting up a database requires access to high-performance compute resources and large disk space.
For example, in November 2019 Google BigQuery's Ethereum database contained more than a terabyte of data.
Moreover, developers still need to manually implement transformation and formatting functionality.
As these databases contain raw data -- e.g., log entries and transaction results are encoded as byte strings -- users also need to implement decoding functions.

By contrast, the proposed framework allows users to only extract the data that is relevant to their use cases.
The previous version of the framework, called BloXES \cite{Klinkmuller+.2019}, was centered around the extraction of data from log entries that were created with Solidity's event API which is the predominant way of logging on Ethereum \cite{Klinkmuller+.2019}.

ELF now allows users to query Ethereum in more flexible ways.
Essentially, the framework retrieves data from Ethereum block by block in historical order, i.e., in the order in which the data was created and included.
Users can then apply the following five types of filters to select relevant entities and their attributes.
The \emph{block filter} allows for selecting blocks whose block number is in the interval $[$\emph{from,to}$]$ and provides access to the attributes of those blocks like mining difficulty or consumed gas.
The \emph{transaction filter} must be nested within a block filter.
It can be used to narrow down the set of transactions within the selected blocks based on account addresses of senders and recipients.
Within the scope of such a filter, 
transaction attributes, such as gas price or transferred value, are accessible.
Similarly, the \emph{log entry filter} enables users to select log entries that were emitted by smart contracts during transaction execution via the event API.
To this end, users need to specify the relevant smart contract addresses and the event signature.
A log entry filter must be combined with a transaction or block filter.
It provides access to log entry attributes and the event signature parameters.
The \emph{smart contract state filter} allows for querying state information of smart contracts.
To this end, developers must specify the contract address and the member variables or functions.
Note that these variables and functions must be part of the contract's public API.
This filter must be nested within a block filter. 
While the first four filters rely on predefined criteria,  
the \emph{generic filter} allows users to introduce arbitrary criteria which can rely on entity or user-defined variables.
For example, users could filter transactions based on the consumed gas.
A generic filter can be nested into any other filter, but does not provide access to new variables.

\capability{Cost-efficient logging}
Writing data into the transaction log consumes \emph{gas} on Ethereum, and is therefore associated with costs \cite{Wood}.
Gas reflects computational and storage demand on a blockchain system, and therefore
developers should consider options for reducing gas consumption. 
ELF offers support to this end in the following ways.
First, it allows for extracting data not only from log entries, but also from blocks, transactions, and smart contract states.
Hence, data obtainable from those entities does not need to be logged.
Second, data transformation operations can be moved to the manifest, making transformations for logging purposes in smart contracts obsolete. 
Lastly, as the gas cost of a log data item is proportional to the size of the value, the generator supports developers in creating compression functionality.
In particular, developers can specify \emph{value dictionaries} and \emph{bit mappings} in the manifest from which the generator then derives custom logging source code that relies on the event API and that can be integrated into a new smart contract.
A value dictionary can be used to define small codes that are logged instead of larger values; during extraction they are decoded.
This can, e.g., be used to log an integer instead of a string, which uses less space and thus incurs a lower gas cost.
Similarly, bit mappings can be used to combine multiple short ranged attributes into a single value via Solidity's bit-level operators.

\capability{Extensible transformation operators}
The framework offers a basic set of data transformation operators. 
The results of these operators can be assigned to user-defined variables.
Additionally, developers can integrate custom operators.
Besides transforming data, those operators could also extend the query functionality.
For example, methods for querying external systems might be added to facilitate the merging of external and blockchain data.
Or non-standardized Ethereum client APIs might be incorporated, such as the transaction replay function offered by Geth or Parity for obtaining more fine-grained data from smart contracts states or transaction executions.

\capability{Dependency updates}
Throughout the life of a dapp, smart contracts might be exchanged to roll out new or improved functionality.
To ensure that a dapp is upgradable, developers can rely on patterns like satellite contracts and contract registers \cite{Woehrer+Zdun.2018}.
This, however, has implications for extracting dapp data, for example, when a smart contract address for which log entry or state data was extracted must be replaced by the address of the newly deployed contract.
In contrast to BloXES, ELF now supports such dynamic updates:
smart contract addresses can be stored in variables, which are updated from queries to contract registers or log entries for such updates. 

\capability{Batch and streaming export}
To support users in analyzing the extracted data, the framework supports the export of entity and user-defined variables into three data formats (textual logs, CSV, and XES) which can be readily loaded into common log analysis tools.
ELF can log multiple formats at the same time, and may even be configured to split data into different files.
Moreover, users can choose between two modes of export.
In \emph{batch} mode, the script runs once and exports all data into (possibly large) files, one file per specified output.
In contrast, the \emph{streaming} mode exports one file per specified output \emph{for each block} on the main chain, and keeps doing so for each new block until terminated.
As such, ELF supports both static log analysis and near-real-time monitoring. 
When using the streaming mode on proof-of-work blockchains, users must be aware that the consensus mechanism can result in multiple alternative blocks with the same block number being proposed by different miners.
If the consensus protocol decides on a different version of a block than the one first received by the local Ethereum node, the block overwrites the first one.
In this case, ELF will output a second file for the same block number, effectively overwriting the previously exported data. 
Note that in the case of the batch mode, the exported files only contain the data that was present in the Ethereum node when data for the last relevant block was extracted. 
\section{IMPLEMENTATION DETAILS}

The framework is implemented in Java v13 and is publicly available \cite{ELF.2019}. 
%
The framework's design is oriented towards guidelines for language implementation \cite{Parr.2009}.
That is, the validator is the central component. 
It relies on the parser generator library ANTLR4 for syntax definition and syntactic parsing.
Semantic analysis is implemented as a set of custom rules on top of the syntax tree returned by ANTLR4.
This amongst others includes checks for type compatibility and correct operator usage.
If there are no syntactical or semantic errors, the syntax tree can be passed to the extractor or generator.

Details regarding the logic for generating logging functionality from bit mappings and value dictionaries are provided in \cite{Klinkmuller+.2019}.
The extractor is designed as a configurable ETL process. 
The integration with nodes on an Ethereum network is based on the Web3j library which provides wrappers for Ethereum's JSON-RPC API.
Moreover, while the export of textual logs and CSV files relies on the Java Class Library, for exporting data according to the XES standard \cite{IEEE.2016} the extractor uses the OpenXES library.
\section{CASE STUDIES}

\begin{figure*}[t]
	\centering
	\begin{subfigure}[t]{.5\textwidth}
		\centering
		\includegraphics[width=.98\textwidth]{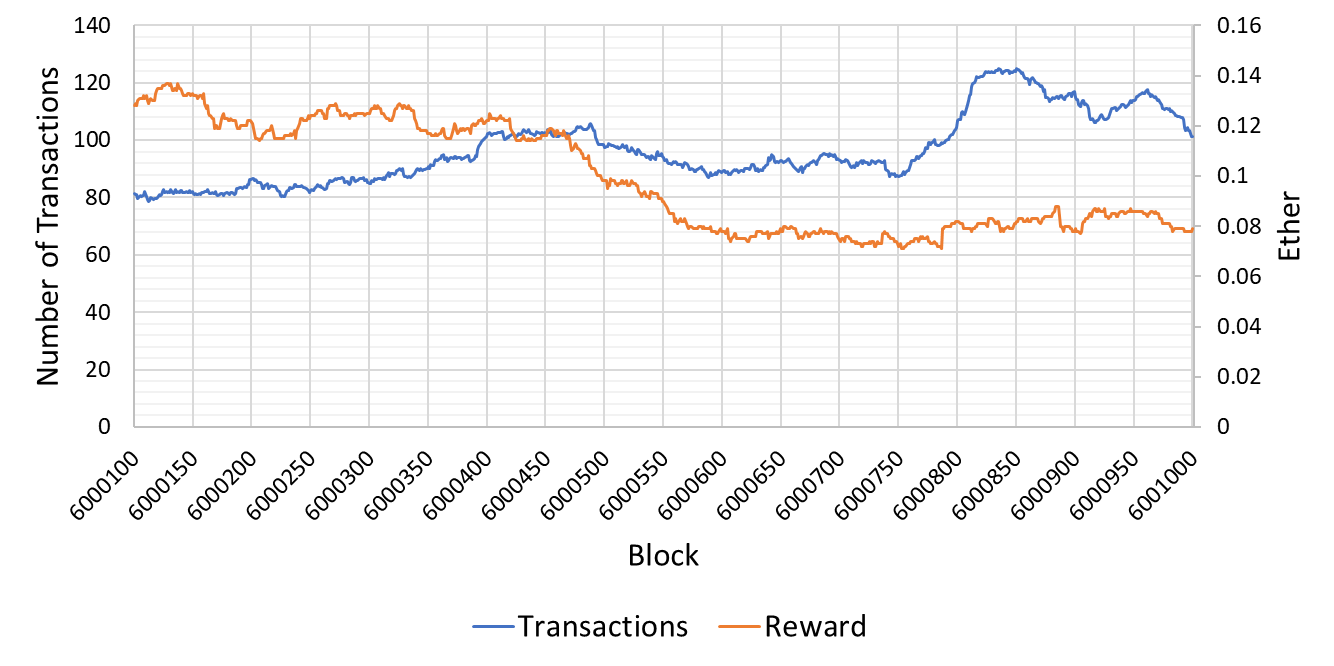}
		\caption{Block statistics}
	\label{fig:casestudies:a}
	\end{subfigure}%
	\begin{subfigure}[t]{.5\textwidth}
		\centering
		\includegraphics[width=.98\textwidth]{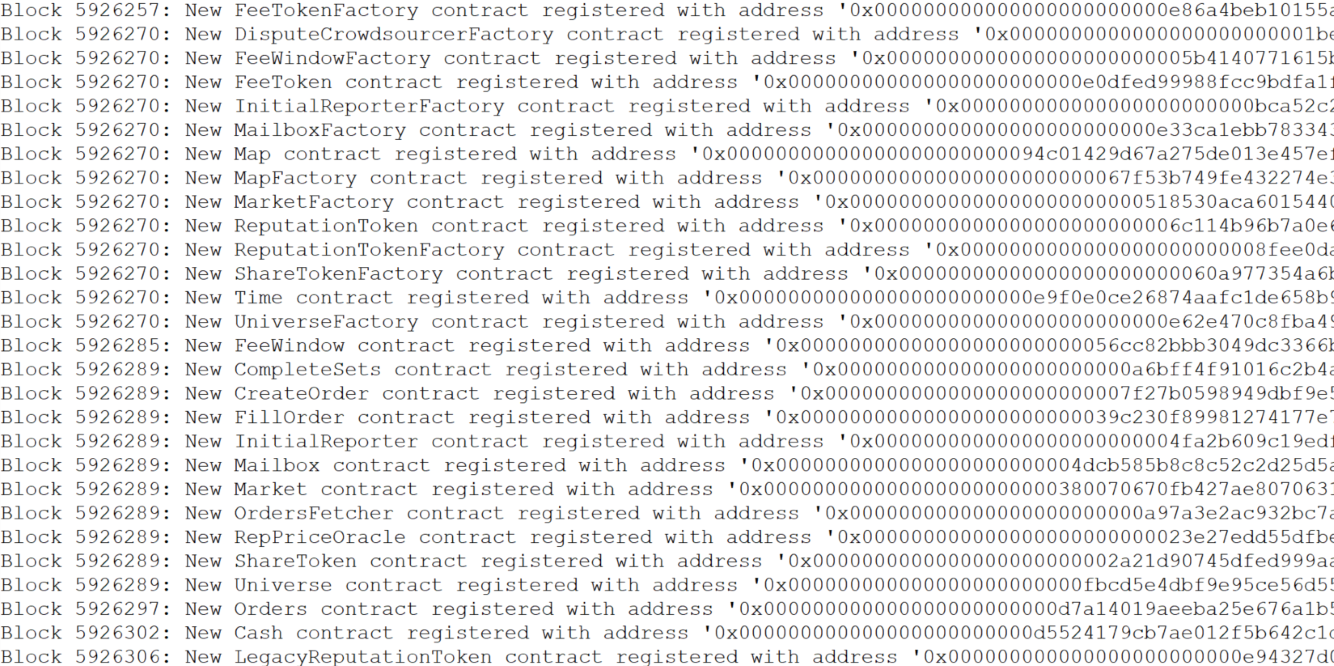}
		\caption{Augur contract register updates}
		\label{fig:casestudies:b}
	\end{subfigure}%
	\\[1em]
	\begin{subfigure}[t]{\textwidth}
		\centering
		\includegraphics[width=.8\textwidth]{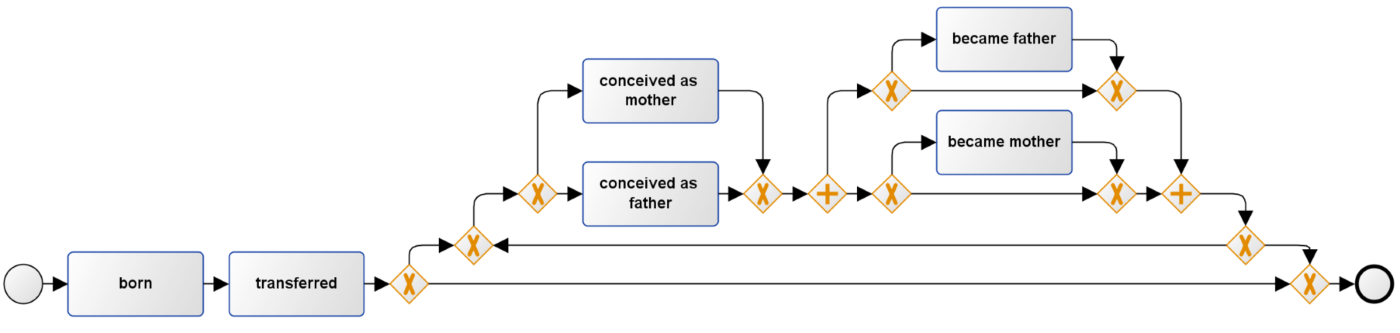}
		\caption{CryptoKitties lifecycle process (notation: BPMN)}
	\label{fig:casestudies:c}
	\end{subfigure}%
	\caption{Case studies}
	\label{fig:casestudies}
\end{figure*}

To demonstrate the applicability of ELF, we conducted three case studies focusing on public Ethereum to allow replication -- ELF is, however, just as applicable for permissioned Ethereum blockchains.
The respective manifests are part of the publicly available source code \cite{ELF.2019}.

The extracted data for all case studies is visualized in \autoref{fig:casestudies}.
First, \autoref{fig:casestudies:a} shows smoothed time series of basic network statistics which are relevant to  network behavior monitoring: the number of included transactions and the reward from transaction inclusion (exclusive of static block rewards). 
Per block, we counted the transactions and summed up the product of consumed gas and gas price for each transaction. 
Finally, the data was exported into a CSV file. 
The respective manifest comprises 14 lines.

Second, we investigated Augur, a decentralized prediction market with a complex architecture centered around a contract register.
We used ELF to setup a listener that monitors Augur's architecture and sends out log messages containing details about contract reference updates. 
\autoref{fig:casestudies:b} shows the updates that occurred during dapp deployment. 
The manifest is 234 lines long.

Third, we analyzed the well-known Ethereum-based game CryptoKitties, in which virtual cats can be bred and traded. 
Using a manifest with 60 lines, we queried information related to the cats' life-cycles from log entries emitted by this dapp. 
We exported the data into an XES file, allowing us to visualize the life-cycle as a process model (\autoref{fig:casestudies:c}) and to confirm that the dapp correctly implements the game rules. 
Note that this case study is essentially replicating the BloXES evaluation from \cite{Klinkmuller+.2019}.
Yet, here we extract a noise-free XES log by only considering cats born within the specified block range.
This was not possible with BloXES. 

\section{SUMMARY}

In this article, we introduced and practically demonstrated ELF, a generic and highly configurable logging framework for Ethereum.
ELF supports a wide-range of log extraction scenarios from Ethereum nodes into different formats, minimizing the amount of required code.
Additionally, ELF assists developers in generating cost-efficient logging code, readily embeddable into dapps. 
The value of the framework has been demonstrated with three case studies that address a variety of logging scenarios, each requiring different framework functionality.
In future work, we plan to improve ELF's run-time efficiency and to generalize it to other blockchain platforms.

\vspace{1em}
\bibliography{references} 
\bibliographystyle{ieeetr}

%
%
%
%

\end{document}